\documentclass[prl,preprintnumbers,aps,twocolumn]{revtex4-2}
\usepackage{amsmath} 
\usepackage{amssymb}
\usepackage{graphicx,epsfig}
\usepackage[mathscr]{eucal}
\usepackage{color}
\usepackage{natbib}
\usepackage[english]{babel}
\usepackage{tabularx}
\usepackage{setspace}
\usepackage{xspace}
\usepackage{multirow}
\usepackage[usenames,dvipsnames]{xcolor}

\def\to{\rightarrow}

\usepackage{booktabs,cellspace}

\usepackage{scalefnt,pstricks}

\def\Wbb{\ensuremath{W b \bar b}\xspace}

\newcommand\Matrix{{\sc Matrix}\xspace}

\newcommand\binone{{\it bin I}\xspace}
\newcommand\bintwo{{\it bin II}\xspace}

\usepackage{array}
\newcolumntype{L}[1]{>{\raggedright\let\newline\\\arraybackslash\hspace{0pt}}m{#1}}
\newcolumntype{C}[1]{>{\centering\let\newline\\\arraybackslash\hspace{0pt}}m{#1}}
\newcolumntype{R}[1]{>{\raggedleft\let\newline\\\arraybackslash\hspace{0pt}}m{#1}}

\usepackage[colorlinks=true,allcolors={blue!70!black}]{hyperref}

\begin{document} 
\preprint{
ZU-TH 60/22,
TIF-UNIMI-2022-22,
PSI-PR-22-35
}

\title{Associated production of a $W$ boson and massive bottom quarks \\at next-to-next-to-leading order in QCD}

\author{Luca Buonocore$^{a}$, Simone Devoto$^{b}$, Stefan Kallweit$^{c}$, Javier Mazzitelli$^{d}$,\\
 Luca Rottoli$^{a}$ and Chiara Savoini$^{a}$\vspace{1em}}

\affiliation{(a) Physik Institut, Universit\"at Z\"urich, CH-8057 Z\"urich, Switzerland,\\
(b) Dipartimento di Fisica, Universit\`a degli Studi di Milano, and INFN, Sezione di Milano, I-20133 Milano, Italy\\
(c) Dipartimento di Fisica, Universit\`{a} degli Studi di Milano-Bicocca and INFN, Sezione di Milano-Bicocca, I-20126, Milan, Italy,\\
(d) Paul Scherrer Institut, CH-5232 Villigen PSI, Switzerland}

\begin{abstract}

We present the first calculation for the hadroproduction of a $W$ boson in association with a massive bottom ($b$)
quark--antiquark pair at next-to-next-to-leading order (NNLO) in QCD perturbation theory.
We exploit the hierarchy between the $b$-quark
mass and the characteristic energy scale of the process to obtain a reliable analytic expression for
the two-loop virtual amplitude with three massive legs, starting from the corresponding result available for massless $b$ quarks. 
The use of massive $b$ quarks avoids the ambiguities associated with the correct flavour assignment in massless
calculations, paving the way to a more realistic comparison with experimental data.
We present phenomenological results considering proton--proton collisions at centre-of-mass
energy \mbox{$\sqrt{s}=13.6$\,TeV} for inclusive $Wb \bar b$ production and within a fiducial region relevant for the
associated production of a $W$ boson and a Higgs boson decaying into a $b \bar b$ pair, for which $Wb \bar b$
production represents one of the most relevant backgrounds. We find that the NNLO corrections
are substantial and that their inclusion is mandatory to obtain reliable predictions.

\end{abstract}

\maketitle

\paragraph{Introduction.}

Many relevant processes at the Large Hadron Collider (LHC) are characterised by signatures featuring one or more bottom ($b$) quarks in the final state.
Among these processes, prominent examples are the production of top quarks, which decay almost exclusively into a $W$ boson and a $b$ quark, and the associated production of a $Z$ or a $W$ boson and a Higgs ($H$) boson decaying into a $b \bar b$ pair.  
Furthermore, the cross section for the production of $B$ hadrons and their decay products constitutes a relevant part of the physics programme at the LHC, both to better characterise the properties of heavy-quark production and to model backgrounds for New-Physics searches.

Bottom quarks cannot be directly observed experimentally, but are detected either at the level of reconstructed $B$ hadrons or by associating a $b$ flavour to final-state jets (i.e.\ collimated bunches of hadrons) which contain one or more $B$ hadrons.
Therefore, a meaningful flavour assignment to the final-state jets is essential to allow for a comparison between theory and data.

From a theoretical point of view such assignment is delicate since it may lead to ill-defined cross sections in higher-order calculations due to the enhanced sensitivity to infrared configurations.
In particular, it is well known that problems related to infrared unsafety arise if the $b$-quark mass $m_b$ is neglected in a theoretical calculation.
Those were first addressed by suitably modifying the $k_t$ jet clustering algorithm~\cite{Catani:1993hr,Ellis:1993tq} for flavoured partons in Ref.~\cite{Banfi:2006hf}, and novel strategies were developed very recently~\cite{Czakon:2022wam,Gauld:2022lem,Caletti:2022hnc,Caletti:2022glq}.
However, jets defined via such approaches generally differ from those identified in the experimental analyses, thus introducing an ambiguity when comparing theoretical predictions with data at the level of reconstructed jets.
In this respect, calculations that fully retain the dependence on the $b$-quark mass are desirable since the quark mass acts as the physical infrared regulator.
This allows the same jet reconstruction algorithm to be used both in the theoretical calculation and in the experimental analysis, thus removing any ambiguity.

In this letter we focus on the associated production of a $W$ boson and a massive bottom quark--antiquark pair, which constitutes one of the main backgrounds to $WH$ and single-top production.
The cross section for the production of a $W$ boson in association with $b$ quarks has been measured both at the Tevatron~\cite{D0:2004prj,D0:2012qt} and the LHC~\cite{ATLAS:2011jbx,ATLAS:2013gjg,CMS:2013xis,CMS:2016eha}.
Next-to-leading-order~(NLO) corrections in QCD for $\Wbb$ production were first computed with massless $b$ quarks in Ref.~\cite{Ellis:1998fv} in the so-called 5-flavour scheme, where the emission of collinear radiation is resummed to all orders into a $b$-quark parton distribution function~(PDF).
A series of studies for massive $b$ quarks~\cite{FebresCordero:2006nvf,FebresCordero:2009xzo}, performed in the so-called 4-flavour scheme, showed that the inclusion of mass effects is phenomenologically relevant.
A consistent combination of the 4- and 5-flavour schemes has been considered in Refs.~\cite{Campbell:2008hh,Campbell:2011udy}, using the matching conditions computed in Ref.~\cite{Cacciari:1998it}.
The NLO corrections were found to be very large, and effects beyond NLO were investigated in a series of works~\cite{Oleari:2011ey,Frederix:2011qg,Luisoni:2015mpa,Anger:2017glm}.

Very recently, a first calculation with massless $b$ quarks was performed at next-to-next-to-leading-order (NNLO)~\cite{Hartanto:2022qhh} using the flavour-$k_t$ algorithm of Ref.~\cite{Banfi:2006hf}.
To reduce the issues related to flavour assignment when comparing with experimental analyses, which widely exploit the anti-$k_T$ algorithm~\cite{Cacciari:2008gp}, its flavour-aware modification of Ref.~\cite{Czakon:2022wam} was used to present phenomenological results in Ref.~\cite{Hartanto:2022ypo}.
Nonetheless, this approach is still not completely satisfactory since intrinsic ambiguities remain when comparing massless predictions with data.
Indeed, the calculation of Ref.~\cite{Hartanto:2022ypo} depends on an unphysical parameter, necessary to make the calculation infrared safe, that must be tuned with data.

In this work we present the first computation of NNLO corrections to $\Wbb$ production with massive $b$ quarks.
Our results are obtained  by retaining the exact dependence on the $b$-quark mass in all the contributions but the two-loop virtual amplitude.
For the latter, we rely on the recently computed results for the massless $b$-quark case in the
leading-colour approximation~\cite{Badger:2021nhg,Abreu:2021asb,Hartanto:2022qhh}. Through a massification procedure~\cite{Penin:2005eh,Mitov:2006xs,Becher:2007cu,Engel:2018fsb}  we consistently include mass effects up to power-suppressed terms of $\mathcal{O}(m_b/Q)$, exploiting the hierarchy of energy scales between $m_b$ and the hard scale $Q \sim m_W$ of the process.

\paragraph{Calculation.}

We start by discussing our calculation for $\Wbb$ production at NNLO.
To deal with infrared singularities arising at intermediate stages of the computation, we use the $q_T$ subtraction formalism~\cite{Catani:2007vq}, which exploits the knowledge of the singular behaviour of the transverse momentum of the produced final state ($q_T$) in the limit \mbox{$q_T \rightarrow 0$}.
The approach was first developed for colour-singlet production and subsequently extended in Refs.~\cite{Bonciani:2015sha,Catani:2019iny,Catani:2019hip,Catani:2020kkl} to deal with massive quarks in the final state.
The extension of the formalism to $W b \bar b$ production does not pose additional complications from a conceptual point of view, but requires the correct treatment of soft QCD radiation in a generic configuration  where the heavy-quark pair is accompanied by a colourless final state~\cite{Catani:202ybbb, Ju:2022wia}, as recently applied to $t\bar t H$ production~\cite{Catani:2021cbl,Catani:2022mfv}.

Schematically, the differential NNLO cross section can be written as
\begin{equation}
d \sigma_{\rm NNLO} = \mathcal H \otimes d\sigma_{\rm LO} + \lim_{r_{\rm cut}\to 0} [d \sigma_{\rm R} - d \sigma_{\rm CT}] _{r > r_{\rm cut}}\,.
\end{equation}
Here $\sigma_{\rm LO}$ is the differential LO cross section, convoluted with the perturbatively computable function $\mathcal H$; the term $d \sigma_{\rm R}$ is the \emph{real} contribution and $d \sigma_{\rm CT}$ the $q_T$ subtraction \emph{counterterm}. 
The real contribution corresponds to a configuration where the $W b \bar b$ final state is accompanied by additional QCD radiation with transverse momentum \mbox{$q_T > 0$}, which is calculable at NLO accuracy using local-subtraction methods~\cite{Frixione:1995ms,Catani:1996vz,Catani:2002hc}.
In the limit \mbox{$q_T \rightarrow 0$} it develops logarithmic divergences which are cancelled by $d \sigma_{\rm CT}$.
The cancellation is performed at small, finite values of the cutoff $r_{\rm cut}$ applied on $r=q_T/M$, where $M$ is the invariant mass of the $Wb\bar b$ final state.
An extrapolation to the limit \mbox{$r_{\rm cut} \rightarrow 0$} is taken~\cite{Grazzini:2017mhc,Buonocore:2021tke}, thereby ensuring that the final result is independent of $r_{\rm cut}$.

The contribution of the massive one- and two-loop virtual amplitudes is part of the coefficient $\mathcal H$, which admits a perturbative expansion in the strong coupling $\alpha_s$\,,
\begin{equation}
\mathcal H \left(\alpha_s \right) = 1 + \alpha_s  \mathcal H_1 + \alpha_s^2 \mathcal H_2  + \ldots  \, .
\end{equation}
In addition to these process-dependent quantities, the coefficient $\mathcal H$ contains universal contributions of soft and collinear origin encoded in soft~\cite{Catani:2014qha,Angeles-Martinez:2018mqh,Catani:2019hip,Catani:202ybbb} and beam functions~\cite{Catani:2012qa,Gehrmann:2014yya,Lubbert:2016rku,Echevarria:2016scs}.
The only missing ingredient to reach NNLO accuracy for $W b \bar b$ with massive $b$ quarks is the two-loop coefficient $\mathcal{M}^{m}_{2}$, which enters the coefficient $\mathcal{H}_{2}$, of the renormalised virtual amplitude,
\begin{equation}\label{eq:massiveampliexp}
\mathcal{M}^{m}(\alpha_s) = \alpha_s (\mathcal{M}^{m}_{0} +  \alpha_s \mathcal{M}^{m}_{1} + \alpha_s^2  \mathcal{M}^{m}_{2} + \ldots )\,,
\end{equation}
where we have factored out a power of $\alpha_s$ since the process starts at this order.
In the following, we describe the massification procedure that allows us to construct a reliable approximation of  $\mathcal{M}^{m}_{2}$ based on the results for the two-loop amplitude with massless $b$ quarks.

Our starting point is the observation that there is a hierarchy between the mass of the $b$ quark $m_b$ and the hard scale $Q$ probed in the process.
As a consequence, the contributions to the virtual amplitude stemming from power corrections $\mathcal O(m_b/Q)$ are phenomenologically negligible, as we explicitly verified in the one-loop case.
This provides a huge simplification for the construction of the massive amplitude since only logarithmically enhanced (powers of $\ln (m_b/Q)$) and constant terms must be considered~\footnote{We note that the logarithmically enhanced terms appear also in real and real-virtual contributions. At LO and at NLO there is an exact cancellation between these logarithmic terms and the result is finite in the $m_b \rightarrow 0$ limit. At NNLO a residual dependence on $\ln (m_b/Q)$ remains, which reflects the necessity of retaining a finite value of $m_b$ lest the calculation become IR unsafe.}.
These terms, dominant in the \mbox{$m_b \rightarrow 0$} limit, are universally related to the singular behaviour of the corresponding massless amplitude by QCD factorisation properties~\cite{Mitov:2006xs}.
As a consequence, we can exploit this connection to map the $\epsilon$ poles of collinear origin, present in the dimensionally regularised massless amplitude, into logarithms of $m_b$ and constant terms to reconstruct the result for the massive amplitude.

The two-loop massive amplitude $\mathcal{M}^{m}_{2}$ can be written as
\begin{equation}\label{eq:massification}
\mathcal{M}^{m}_{2} = \mathcal{M}^{m=0}_{2} + Z_{[q]}^{1} \mathcal{M}^{m=0}_{1} + Z_{[q]}^{2} \mathcal{M}^{m=0}_{0}\, ,
\end{equation}
where $\mathcal{M}^{m=0}_{k}$ are the renormalised coefficients of the massless amplitude~\cite{Badger:2021nhg,Abreu:2021asb,Hartanto:2022qhh}, in analogy with Eq.~\eqref{eq:massiveampliexp}, and $Z^{k}_{[q]} $ are the perturbative coefficients of the massification form factor for the quark case, computed in Ref.~\cite{Mitov:2006xs}.
We note that all the quantities appearing in Eq.~\eqref{eq:massification} contain infrared poles and, thus, the knowledge of $\mathcal{M}^{m=0}_{1}$ and $Z_{[q]}^{1} $ up to $\mathcal{O}(\epsilon^2)$ is mandatory.
As a strong check of our construction we have verified that the resulting infrared structure of $\mathcal{M}^{m}_{2}$ is consistent with that predicted in Ref.~\cite{Ferroglia:2009ep}.

We further note that the massless two-loop amplitude $\mathcal{M}^{m=0}_{2}$ of Refs.~\cite{Badger:2021nhg,Abreu:2021asb,Hartanto:2022qhh} is available at leading colour whereas all the other terms in Eq.~\eqref{eq:massification} are known in full colour (with the only exception of the $\mathcal{O}(\epsilon^k)$, \mbox{$k\geq1$} terms of $\mathcal{M}^{m=0}_{1}$).
In our massification procedure we retain the full colour dependence in all the known terms.

We perform the calculation within the \Matrix framework~\cite{Grazzini:2017mhc}, suitably extended to $W b \bar b$ production.
The evaluation of tree-level and one-loop matrix elements with massive $b$ quarks is performed via the \textsc{OpenLoops}~\cite{Cascioli:2011va,Buccioni:2017yxi,Buccioni:2019sur} and \textsc{Recola}~\cite{Actis:2012qn,Actis:2016mpe,Denner:2017wsf,Denner:2016kdg} codes.
In our calculation we do not consider diagrams with massive-quark loops in the real--virtual contributions since analogous diagrams appearing in the two-loop amplitude are not recovered by the massification procedure described above.
Accordingly, we do not include the real subprocess with four massive $b$ quarks entering at NNLO.
We have verified that the latter contribution, which constitutes an estimate of the impact of the neglected diagrams, has a negligible effect in our results.
The contribution of the two-loop virtual amplitude with massive $b$ quarks is computed in a dedicated C++ code~\cite{github} which is interfaced to \Matrix and uses \textsc{OpenLoops}, the \textsc{PentagonFunctions-cpp}~\cite{Chicherin:2021dyp} package and the amplitudes from Ref.~\cite{Abreu:2021asb}.

\paragraph{Phenomenology.}

We now move on to discussing the phenomenological impact of the NNLO corrections.
We consider proton--proton collisions at the centre-of-mass energy \mbox{$\sqrt{s}=13.6$\,TeV} and focus on two different setups: i) $W b \bar b$ production in the inclusive phase space with no fiducial cuts applied and ii) $W b \bar b$ production within a fiducial region that closely resembles that considered in the ATLAS analysis of Ref.~\cite{ATLAS:2020fcp}, which is relevant for the associated production of a $W$ boson and a Higgs boson decaying into a $b \bar b$ pair.
We compute the full process \mbox{$p p \rightarrow e^+ \nu_e b \bar b$}.
A comparison with the results of Refs.~\cite{Hartanto:2022qhh,Hartanto:2022ypo}, obtained with massless $b$ quarks, is presented in the Appendix.
We find that the two results overlap within the respective uncertainty bands and that our systematic error is smaller than the ambiguities related to the use of the flavour anti-$k_t$ algorithm. In particular, we find our result very stable under a conservative variation of the physical value of the $b$-quark mass, whereas that of Ref.~\cite{Hartanto:2022ypo} is subject to a much larger uncertainty upon variation of the unphysical parameter characterising the flavour anti-$k_T$ algorithm.

We use NNPDF3.0 LO PDFs~\cite{Ball:2012cx} with \mbox{$\alpha_s=0.118$} for LO predictions and NNPDF3.1 PDFs~\cite{NNPDF:2017mvq} NLO (NNLO) PDFs with \mbox{$\alpha_s=0.118$} for those at NLO (NNLO) through the LHAPDF interface~\cite{Buckley:2014ana}.
We consistently adopt parton densities with \mbox{$n_f=4$} for our calculation.
We employ the $G_\mu$ scheme with complex masses~\cite{Denner:1999gp} and the following electroweak (EW) input parameters:
$G_F = 1.16638 \times 10^{-5}$\,GeV$^{-2}$, $m_W = 80.379$\,GeV, $\Gamma_W = 2.0855$\,GeV, $m_Z = 91.1876$\,GeV, $\Gamma_Z = 2.4952$\,GeV.
We use a diagonal CKM matrix.
We set the heavy-quark masses to \mbox{$m_t=173.3$\,GeV} and \mbox{$m_b=4.92$\,GeV} (if not stated otherwise), the latter corresponding to the value chosen in the NNPDF3.1 analysis.

We start our discussion with inclusive $W b \bar b$ production.
We note that this quantity is well-defined in a scheme with massive $b$ quarks, while it cannot be computed in a massless calculation (even at LO) due to the unregularised \mbox{$g \rightarrow b \bar b$} splitting.
Although experimentally the inclusive cross section for $W b \bar b$ production is not directly measurable, the results for the inclusive case are useful to assess the impact of NNLO corrections and to investigate the convergence properties of the perturbative series.
In addition, the possibility to compute theoretical predictions in the whole physically accessible phase space makes our computation particularly suitable for developing a fully-fledged Monte Carlo event generator.
This could be achieved by matching our NNLO results to parton showers, for instance through the MiNNLO$_{\rm PS}$ method developed in Refs.~\cite{Monni:2019whf,Monni:2020nks} and extended to final-state heavy quarks in Ref.~\cite{Mazzitelli:2020jio}.

Our results are collected in Tab.~\ref{tab:xs} where we show the cross section at LO, NLO and NNLO. We set the central renormalisation and factorisation scales \mbox{$\mu_R=\mu_F= m_W/2 + m_b$}.
The uncertainty stemming from missing higher-order terms is estimated through variation of $\mu_R$ and $\mu_F$ by a factor of 2 around the central value, whilst keeping \mbox{$1/2 \leq \mu_R/\mu_F \leq 2$}, i.e.\ we use the standard seven-point scale variation band.
We observe that NNLO corrections are necessary for a satisfactory description of this process, since first evidence of perturbative convergence appears at this order.
The NLO result is more than 3 times larger than the LO prediction, and the LO scale uncertainties vastly underestimate the effects from missing higher orders.
NLO corrections were found to be large already in previous calculations, due to the quark--gluon channel opening up only at NLO, which also explains that no reduction of scale uncertainties is observed when going from LO to NLO.

\begin{table}[t]
\centering
\renewcommand{\arraystretch}{1.5}
\setlength{\tabcolsep}{0.4em}
\begin{tabular}{clll}
  order &
  \multicolumn{1}{c}{$\sigma_{\rm incl}\, [{\rm pb}]$} &
  \multicolumn{1}{c}{$\sigma_{\rm fid}^{\textnormal{\binone}}\, [{\rm fb}]$} &
  \multicolumn{1}{c}{$\sigma_{\rm fid}^{\textnormal{\bintwo}}\, [{\rm fb}]$}
  \\
\hline
  LO &
  $ 18.270(2)^{+28\%}_{-20\%}$ &
  $ \phantom{0}35.49(1)^{+25\%}_{-18\%}$ &
  $ \phantom{0}8.627(1)^{+25\%}_{-18\%}$
  \\
  NLO &
  $ 60.851(7)^{+31\%}_{-21\%}$ &
  $ 137.20(5)^{+34\%}_{-23\%}$ &
  $ 37.24(1)^{+38\%}_{-24\%}$
  \\
  NNLO &
  $95.54(8)^{+21\%}_{-17\%}$ &
  $199.5(8)^{+17\%}_{-15\%}$ &
  $56.34(8)^{+19\%}_{-17\%}$
  \\
\end{tabular}
\caption{\label{tab:xs}Inclusive and fiducial cross sections for $W b \bar b$ production at different perturbative orders. The numbers in parenthesis indicate the statistical uncertainties of our results. At NNLO, the error also includes the extrapolation uncertainty.}
\end{table}

The NNLO corrections are sizable and increase the NLO result by more than 50$\%$, but with reduced scale uncertainties.
The NLO and NNLO uncertainty bands, however, partially overlap, indicating \ that the series is
starting to converge. This is supported by the fact that all partonic channels are eventually open at NNLO.

\begin{figure*}[t]
  \includegraphics[width=0.48\textwidth]{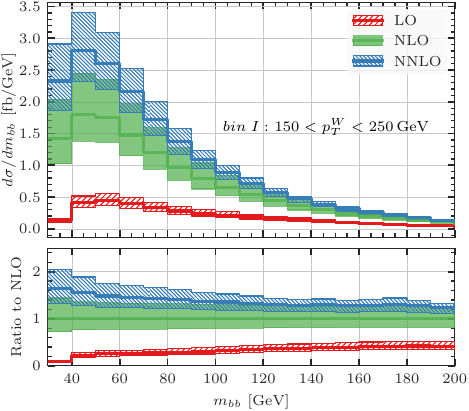}\hfill
  \includegraphics[width=0.48\textwidth]{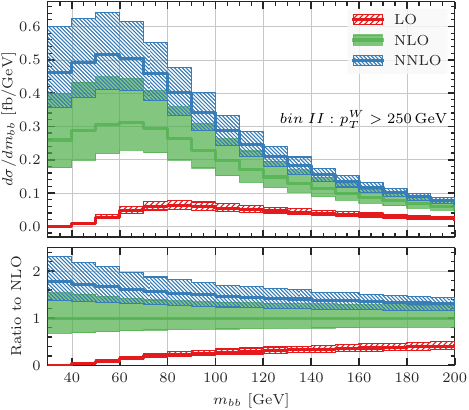}
  \caption{Invariant-mass distribution of the $b$-jet pair in fiducial region, in \binone (left) and \bintwo (right), at LO (red), NLO (green) and NNLO (blue). See text for more details.}
    \label{fig:wbbfid}
\end{figure*}

We now discuss the results obtained by applying selection cuts to $W b \bar b$ production.
The fiducial region is defined as follows.
We consider jets reconstructed using the anti-$k_T$ algorithm with a jet radius \mbox{$R=0.4$}.
The jets are required to have a minimum transverse momentum of $20$ ($30$)\,GeV if the pseudo-rapidity is \mbox{$|\eta_j| < 2.5$} (\mbox{$2.5 < |\eta_j| < 4.5$}).
We demand exactly two tagged $b$ jets.
The leading $b$ jet is required to have a minimum transverse momentum of $45$\,GeV, and we impose a separation \mbox{$0.5 < \Delta R_{bb} < 2$} between the two $b$ jets.
No veto is applied on additional jets.
The charged lepton is required to have a minimum transverse momentum of $25$\,GeV and \mbox{$|\eta_\ell|<2.5$}.
The fiducial region is partitioned into two bins according to the transverse momentum of the reconstructed $W$ boson: \mbox{$150 < p_T^W < 250 $\,GeV} (\binone); \mbox{$p_T^W > 250 $\,GeV} (\bintwo).

We collect the results for the fiducial cross sections in Tab.~\ref{tab:xs}.
We set the central renormalisation and factorisation scales to \mbox{$\mu_R = \mu_F =
\sqrt{H_T m_{bb}}/2$ }, where $m_{bb}$  is the invariant mass of the $b$-jet pair and $H_T=\sqrt{m_{\ell \nu}^2+p_{T,\ell \nu}^2}+p_{T,b_1}+p_{T,b_2}$, with $p_{T,b_{i}}$ $(i=1,2)$ the transverse momenta of the two $b$ jets. We compute scale uncertainties as in the inclusive case.
The choice of the scale is motivated by the observation that there are two characteristic scales in the fiducial setup: a hard scale  of $\mathcal{O}(H_T)$ and a lower scale of $\mathcal{O}(m_{bb})$ associated with the gluon splitting to the $b\bar b$ pair. A geometric mean effectively takes into account both configurations.
We observe a similar pattern of the higher-order corrections as in the inclusive case.
The NLO corrections are substantial and again way outside the uncertainty bands of the LO result, even more evidently in \bintwo.
The inclusion of NNLO corrections is mandatory, given the poor behaviour of the perturbative expansion.
They turn out to be large both in \binone and especially in \bintwo, where they amount to more than $50\%$ of the NLO result.
The scale uncertainties are reduced from NLO to NNLO, where they are at most $20\%$.

In Fig.~\ref{fig:wbbfid} we show the invariant-mass distribution $m_{bb}$ of the $b$-jet pair in the two bins.
We observe that both the NLO and the NNLO corrections are not uniform in $m_{bb}$, being larger for smaller invariant-mass values. 
As a consequence, the peaks of the differential cross sections are shifted towards lower values of $m_{bb}$ when higher orders are included, to \mbox{$\sim 50$\,GeV} and \mbox{$\sim 70$\,GeV} in \binone and \bintwo, respectively, at NNLO. The peak in \bintwo turns out to be significantly broader than in \binone.
The missing higher-order uncertainty estimated through scale variations is smaller at NNLO than at NLO, with the two bands partially overlapping.

Lastly, we assess the impact of various systematic uncertainties on our central prediction.
We start by considering the uncertainty due to the value of the $b$-quark mass by conservatively lowering its value to \mbox{$m_b=4.2$\,GeV}. Starting at NNLO, such variation probes the additional dependence of $\ln m_b/Q $ associated with the use of a flavour unaware jet clustering algorithm, besides that associated with the use of a 4FS scheme.
The cross section at NNLO  increases by 2\%, which is well within the quoted scale uncertainties, supporting the validity of a fixed-order treatment.
The impact of the massification procedure can be assessed at NLO, by comparing the exact result quoted in Tab.~\ref{tab:xs} with a result obtained via the massified version of the (massless) one-loop amplitude.
We find that the difference amounts to 3\% of the NLO correction, again vastly smaller than the scale uncertainties, thereby providing a strong indication for the reliability of our procedure at NNLO.
Finally, we find that the contribution of the two-loop virtual amplitude computed at leading colour amounts to 2\% of the total NNLO cross section.
Since the leading-colour approximation is expected to reproduce full-colour results within $10\%$ accuracy, we estimate that the error resulting from relying on such approximation is below the percent level on our final results.

\paragraph{Summary and outlook.}

In this letter we have presented the first NNLO calculation for $W b \bar b$ hadroproduction with massive bottom quarks.
Our result is not subject to flavour-tagging ambiguities, which affect calculations with massless bottom quarks starting at the NNLO level, and can be compared directly with experimental data.
Our predictions are obtained with the $q_T$ subtraction formalism, properly extended to this class of processes by evaluating the relevant  soft contributions.
The two-loop amplitude is computed through a massification procedure.
Our result indicates that this is a powerful strategy for obtaining two-loop amplitudes with several massive legs, whose exact computation is beyond the present state of the art.
 
We have presented phenomenological results at the LHC, both at the inclusive level and within fiducial cuts.
We found that the NNLO corrections are sizeable and that the perturbative series starts manifesting convergence properties only if these are included.
The inclusion of NNLO corrections would help to alleviate the tension observed in a recent CMS analysis~\cite{CMS:2018nsn}, which needs to apply large corrections factors on top of NLO results to properly estimate the $W b \bar b$ background in the phase space relevant for Higgs production in association with a $W$ boson~\cite{Tricoli:2020uxr}.

Our novel calculation paves the way for several further high-impact applications.
Firstly, it is particularly suitable to be matched to parton showers to construct a fully-fledged Monte Carlo event generator for $W b \bar b$ production.
This will be of paramount importance for the Higgs precision programme at the LHC since $W b \bar b$ constitutes one of the main backgrounds for $WH$ associated production.
Secondly, we foresee an application in $W$+charm production, which is relevant for the determination of the strange-quark PDF.
Finally, the methodology discussed in this paper can be applied to other relevant processes at the LHC, like the production of a $Z$ boson in association with a heavy-quark pair, once the corresponding massless amplitude will become available.

%

\paragraph{Acknowledgements.}
We thank Samuel Abreu, Massimiliano Grazzini and Fabio Maltoni for discussions and comments on the manuscript. We thank Vasily Sotnikov for discussions about the impact of the leading-colour approximation and for a point-wise comparison of the code used for our predictions and a private implementation of the two-loop massive amplitude.
We are grateful to Jonas Lindert for support with the use of OpenLoops amplitudes.
This work is supported by the Swiss National Science Foundation contracts 200020$\_$188464 and PZ00P2$\_$201878, by the UZH Forschungskredit Grants K-72324-03, FK-21-102 and FK-22-099, by the grant PRIN201719AVICI 01, by the ERC Starting Grant 714788 REINVENT, and by the Deutsche Forschungsgemeinschaft under Germany's Excellence Strategy - EXC-2094 - 390783311.

\appendix

\section{Appendix: Comparison to a 5FS calculation}

\begin{table*}
\centering
\renewcommand{\arraystretch}{1.5}
\setlength{\tabcolsep}{0.6em}
\begin{tabular}{cllllll}
  order & \hspace*{1.2em} &
  \multicolumn{1}{c}{$\sigma^{\mathrm{4FS}}$\,[fb]} & \hspace*{1.2em} &
  \multicolumn{1}{c}{$\sigma^{\mathrm{5FS}}_{a=0.05}$\,[fb]} &
  \multicolumn{1}{c}{$\sigma^{\mathrm{5FS}}_{a=0.1}$\,[fb]} &
  \multicolumn{1}{c}{$\sigma^{\mathrm{5FS}}_{a=0.2}$\,[fb]} 
  \\
  \hline
  LO & &
  $210.42(2)^{+21.4\%}_{-16.2\%}$ & &
  $262.52(10)^{+21.4\%}_{-16.1\%}$ &
  $262.47(10)^{+21.4\%}_{-16.1\%}$ &
  $261.71(10)^{+21.4\%}_{-16.1\%}$ 
  \\
  NLO & &
  $ 468.01(5)^{+17.8\%}_{-13.8\%}$ & &
  $ 500.9(8)^{+16.1\%}_{-12.8\%}$ &
  $ 497.8(8)^{+16.0\%}_{-12.7\%}$ &
  $ 486.3(8)^{+15.5\%}_{-12.5\%}$ 
  \\
  NNLO & &
  $652.8(1.6)^{+12.8\%}_{-11.0\%}$& &
  $690(7)^{+10.9\%}_{-9.7\%}$ &
  $677(7)^{+10.4\%}_{-9.4\%}$ &
  $647(7)^{+9.5\%}_{-9.4\%}$
  \\
\end{tabular}
\caption{\label{tab:xs2} Cross sections for $Wb \bar {b}$ production in the 4FS using anti-$k_t$ algorithm and in the 5FS using flavour anti-$k_t$ algorithm with different values of the parameter $a$ (see text for details). The 5FS results are taken from Ref.~\cite{Hartanto:2022ypo}.}
\end{table*}

In this Appendix we compare our result for $Wb\bar b$ production, obtained in the 4-flavour scheme~(4FS) with massive bottom quarks, to the result of Ref.~\cite{Hartanto:2022ypo} computed in the 5-flavour scheme~(5FS) with massless bottom quarks.
For this comparison we set the electroweak parameters to the values used for our predictions in the main text, matching the settings of Ref.~\cite{Hartanto:2022ypo}.
Concerning the input parton densities, both computations use NNPDF3.1 PDF sets~\cite{NNPDF:2017mvq} for NLO and NNLO predictions with $n_f=4$ and $n_f=5$, respectively.
At LO, the massless computation of Ref.~\cite{Hartanto:2022ypo} employs the NNPDF3.1 PDF set, whereas our results are obtained by using NNPDF3.0 PDFs~\cite{Ball:2012cx} with $n_f=4$, since NNPDF3.1 LO PDFs with $n_f=4$ are not available. 
We set $\mu_R=\mu_F=H_T$ as in Ref.~\cite{Hartanto:2022ypo}.

We consider proton--proton collisions at a centre-of-mass energy \mbox{$\sqrt{s}=8$\,TeV}.
The fiducial region is defined by requiring the presence of a charged lepton with \mbox{$p_T^\ell > 30$\,GeV} and \mbox{$|\eta^\ell| < 2.1 $}  and at least two $b$ jets with \mbox{$p_T^b > 25$\,GeV} and \mbox{$|\eta^b| < 2.4 $}.
In the 4FS calculation the $b$ jets are defined via the standard anti-$k_t$ algorithm~\cite{Cacciari:2008gp} with $R=0.5$. Specifically, we assign a $b$ flavour to each jet that contains at least one $b$ or $\bar b$ quark.
On the other hand, in the 5FS calculation the results are computed using the recently proposed flavour anti-$k_t$ algorithm~\cite{Czakon:2022wam} with the same jet radius.
In the latter case, the results depend additionally on the parameter $a$ which acts as a regulator of the infrared divergences.
The anti-$k_t$ algorithm, which is infrared unsafe starting at NNLO in a 5FS calculation, is recovered in the limit \mbox{$a\rightarrow 0$}.
Starting from NNLO, the value of $a$ should be carefully tuned to be sufficiently small to allow for a comparison with experimental data where the anti-$k_t$ clustering algorithm is typically used, yet sufficiently large lest the perturbative convergence of the calculation be spoiled by large logarithms of $a$.
In Ref.~\cite{Hartanto:2022ypo} three values of the parameter $a$ are considered, namely $0.05$, $0.1$ and $0.2$.

In Tab.~\ref{tab:xs2} we compare the LO, NLO and NNLO predictions in the 4FS to the 5FS results for different values of the parameter $a$.
We observe a good agreement between the two calculations within scale uncertainties, with the 4FS being systematically below the 5FS result.
At LO the 4FS is about $20\%$ smaller than the 5FS result, which at this order is essentially independent of $a$.
At NLO and NNLO the difference between the two schemes reduces to below the $10\%$ level, and becomes smaller for larger values of the parameter $a$.
At all perturbative orders the scale variation bands are of the same size in the two computations.

The difference at the level of the central values can be reduced by performing a change of scheme in our computation and by using the same PDFs and strong running coupling of the 5FS calculation.
We have performed this exercise at NLO.
At this order, we can directly use PDFs with \mbox{$n_f=5$} in our 4FS calculation: differences between the two schemes will only start at NNLO since we do not have gluon initiated partonic processes at LO. 
On the other hand, in replacing the strong coupling constant renormalised considering \mbox{$n_f=4$} active flavours with that renormalised with \mbox{$n_f=5$}, we need to take into account the NLO correction~\cite{Cacciari:1998it},
\begin{equation}
\alpha_s^{n_f=4}(\mu) = \alpha_s^{n_f=5}(\mu)\left[1-\alpha_s^{n_f=5}(\mu)\frac{T_R}{3\pi}\ln\frac{\mu^2}{m_b^2}+\ldots\right]\,.
\end{equation}
We find that the NLO result increases up to 481\,fb, which is indeed closer to the 5FS result.
As a further check, we have taken the massless limit of our NLO calculation by performing an extrapolation from results obtained with progressively smaller values of the $b$-quark mass. By this procedure, we find that the result further increases to 491\,fb.
We estimate therefore that the size of the mass corrections at NLO is as large as the impact of the change of scheme.
We notice that the inclusion of higher-order corrections should also reduce differences between the 4FS and 5FS as the two schemes are formally equivalent in all-order QCD.
Since the 4FS computation is sensitive to the value of the $b$-quark mass, we conservatively vary its value down to $4.2$\,GeV.
We find that the NNLO cross section is rather stable upon such variation, showing a marginal \mbox{$\sim 2\%$} increase.
In comparison, the NNLO result in the 5FS features a more pronounced dependence on the values of $a$, where we observe that the predictions for \mbox{$a=0.2$} and \mbox{$a=0.05$} differ by almost $7\%$.

\begin{figure}[t]
  \includegraphics[width=0.48\textwidth]{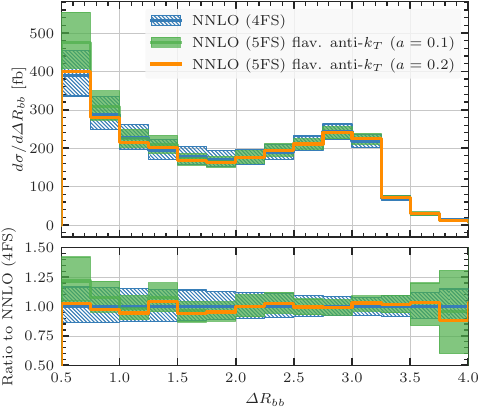}
    \caption{Rapidity and azimuthal angle separation $\Delta R_{bb}$ between the two leading $b$ jets in the 4FS (blue) and in the 5FS with \mbox{$a=0.1$} (green) calculations. For reference, the 5FS result with \mbox{$a=0.2$} is also shown (orange). The 5FS results are taken from Ref.~\cite{Hartanto:2022ypo}.}
    \label{fig:wbb5fs}
\end{figure}

Finally, in Fig.~\ref{fig:wbb5fs} we compare the 4FS results against those obtained in the 5FS with \mbox{$a=0.1$} for the separation $\Delta R_{bb}$ in the rapidity and azimuthal angle between the two leading $b$ jets.
This distribution was shown in Ref.~\cite{Hartanto:2022ypo} to clearly discriminate the flavour anti-$k_t$ clustering algorithm from the flavour-$k_t$ algorithm, as the latter features an enhanced suppression at small values of $\Delta R_{bb}$.
We observe an overall good agreement between the results of the two computations across the whole range of the plot.
The ratio between the two results is largely flat for \mbox{$\Delta R_{bb} \gtrsim 1$}.
For smaller values of $\Delta R_{bb}$, the 5FS result tends to slightly overshoot the 4FS result and gets up to 25\% larger.
On the other hand, the 5FS result with \mbox{$a=0.2$} seems to be in better agreement with our result, both at large and at small values of $\Delta R_{bb}$.

In conclusion, we observe an overall good agreement between the two computations at NNLO, with the 4FS result being \mbox{$\sim 10\%$} smaller than the 5FS result and with scale uncertainties largely overlapping.
We find that the 4FS is largely independent of the value of the $b$-quark mass and tends to be in better agreement with 5FS results obtained with a flavour anti-$k_T$ algorithm if the tuneable $a$ parameter is $\gtrsim 0.1$.
The agreement improves if a change of scheme is applied to the 4FS computation to use the same PDFs and strong coupling as in the 5FS computation.

\bibliography{biblio}

\end{document}